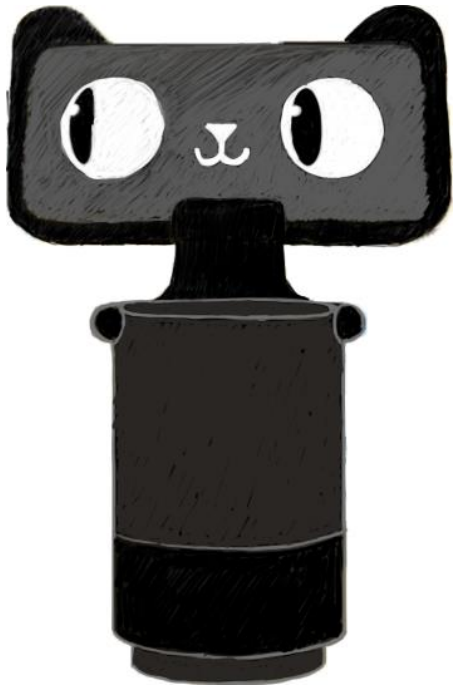

**Figure 1.** Alibaba's smart speaker, Tmall Genie X1, with the Intelligent Personal Assistant (IPA) "AliGenie". A smartphone can be connected to the smart speaker. The front-facing camera is used for visual recognition while the screen displays anthropomorphic and zoomorphic characteristics.

# Studying Breakdowns in Interactions with Smart Speakers

**Mirzel Avdic**
Department of Computer Science
Aarhus University, Denmark
miavd18@cs.au.dk

**Jo Vermeulen**
Department of Computer Science
Aarhus University, Denmark
jo.vermeulen@cs.au.dk

## ABSTRACT

The popularity of voice-controlled smart speakers with intelligent personal assistants (IPAs) like the Amazon Echo and their increasing use as an interface for other Internet of Things (IoT) technologies in the home provides opportunities to study smart speakers as an emerging and ubiquitous IoT device. Prior research has studied how smart speaker usage has unfolded in homes and how the devices have been integrated into people's daily routines. In this context, researchers have also observed instances of smart speakers' 'black box' behaviour. In this position paper, we present findings from a study we conducted to specifically investigate such challenges people experience around intelligibility and control of their smart speakers, for instance, when the smart speaker interfaces with other IoT systems. Reflecting on our findings, we discuss new possible directions for smart speakers including physical intelligibility, situational physical interaction, and providing access to alternative interpretations in shared and non-shared contexts.

## THE TROUBLE WITH SENSING SYSTEMS

Systems that make sense of the environment and act on information they infer have long been discussed by researchers under the name "context-aware systems" [6]; smart speakers and other IoT devices are a recent rendition of those systems. In the early 2000s, researchers voiced concerns about context-aware systems' lack of intelligibility and control [4,8]; i.e. informing users of what the system infers, how it infers this, what it does with that information, and how people can make decisions to take control over the system based on this information [4]. Bellotti et al. [2] highlighted design challenges regarding sensing systems and proposed five questions for designers:

- *Address*: How do I address one (or more) of many possible devices?
- *Attention*: How do I know the system is ready and attending to my actions?
- *Action*: How do I effect a meaningful action, control its extent and possible specify a target or targets for my action?
- *Alignment*: How do I know the system is doing (has done) the right thing?
- *Accident*: How do I avoid mistakes?

These questions have served as inspiration for our study on issues people experience with respect to intelligibility and control of smart speakers as described in 'Our Recent Work' section.

## INTRODUCTION

Given the fact that connected IoT devices are using sophisticated processing of sensor data and are increasingly acting on our behalf [1], it is paramount to develop techniques to extend human control over technological environments and to empower users to better understand the technologies they use.

Smart speakers are one example of these devices. Their popularity presents an opportunity to study potential issues with intelligibility [4] with a common and increasingly connected IoT device, as described in the side bar; some of which Porcheron et al. have observed and referred to as 'black box' behaviour [15]. In their study, participants had difficulties using the smart speaker due to a lack of *interactional resources*, i.e. a lack of clear and informative responses by the smart speaker did not lead the participants in a fruitful direction to recover from breakdowns. This leaves unresolved questions about users' mental model of how a smart speaker works, in what ways users address their smart speaker, and how users recover from mistakes and system breakdowns. While it is unclear whether smart speakers will become fully autonomous, they have the potential to shift toward becoming Autonomous IoT (A-IoT) [1] in homes and may further influence users' lives. It then becomes important to give users possibility to adjust autonomy as well as question the systems' decision making processes in conjunction with users' ongoing task [20]. Researchers have also demonstrated ways to provide intelligibility in other systems [5,6,11–13,18,19], emphasizing textual and visual representations, with some exceptions such as shape change (e.g. [14]).

Porcheron et al. also found that participants in households collaboratively attempt to recover from smart speaker breakdowns, although not always successfully [15]. In a different study, Porcheron et al. found how voice user interfaces democratize smartphones by allowing others at cafés, such as friends, to engage with others' smartphones via voice commands during breakdowns without invitation [17]. These findings show that people tend to collaborate in recovering and using voice controlled devices, yet, it remains unclear to which extent this democratization is socially acceptable with respect to guests in homes with smart speakers present. While Lau et al. [10] recently observed that some users placed their smart speaker in central locations such as living rooms and introduced their device to guests, it is still unknown whether smart speakers are used differently with visitors present in contrast to when users are alone or with family, and if so, how?

From the above, we decided to conduct a study to investigate the following: (1) what are users' mental model of the smart speaker, (2) users addressing their smart speaker, (3) users' recovery from mistakes and system breakdowns, and (4) users' interactions with smart speakers in shared contexts. We discuss our findings and from these we extract directions for future research.

**RESEARCH QUESTIONS**

(1) What are users' mental models of smart speakers?
(2) How do users address their smart speakers?
(3) How do users recover from mistakes and system breakdowns?
(4) How do users use their smart speaker with others in households and/or when having visitors?

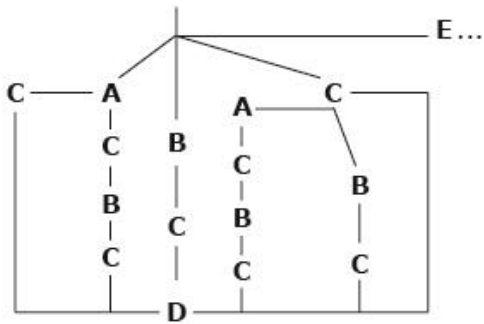

Figure 2. Possible strategies users tend to use to recover from breakdowns.
(A) Orient themselves towards the smart speaker.
(B) Walk up to the smart speaker (facing it).
(C) Retry the request 'X' amount of times.
(D) Try to find the issue on smartphone, use an app to complete the task, or give up entirely.
(E) Other approaches.

## OUR RECENT WORK

To better understand issues people face with smart speakers in terms of 'black box' behaviour, we conducted a study to investigate when users encounter unintelligible behaviour and the strategies they use to recover from it [2]. Our study consisted of an online survey with 117 respondents and 12 semi-structured interviews with smart speaker owners. In the survey, we asked participants to perform specific tasks with their smart speaker, and reflect on their experiences. In the interviews, we primarily steered the conversation around the participant's mental model, experiences with breakdowns, and using the smart speaker as a shared device. Our survey and interviews focused on four research questions (see the side bar). Research questions 2 and 3 were directly inspired by Bellotti et al.'s issues *address* and *accident* [3].

The following are findings from our study with respect to these research questions:

**(1) Mental model**. The 12 interviewees had four main explanations for how their smart speaker functioned; perceiving the device as just a 'dumb' speaker, believing it was action-triggered, thinking that most of the functionality happens in the cloud, and finally, explaining many details in the speech recognition and natural language understanding process step-by-step, which we observed when participants had a technical background.

**(2) Addressing the smart speaker**. In both the online survey and interviews, participants preferred to address the smart speaker using voice alone to keep their hands free. Some reported situational physical interactions with the smart speakers depending on the participant's proximity to the device and circumstances they were in. For instance, one participant physically interacted with the smart speaker when family members would go to bed. To avoid waking them up, the participant utilized the buttons on the smart speaker to trigger it and proceed to whisper a request. Pressing the button was a confirmation to the participant that the smart speaker listened and he then hoped that the device would pick up the request. While the smart speaker's primary interface is voice, it heavily relies on the smartphone for installation, configuration, information, and alternative controls. One participant pointed out that the smartphone is sufficient for relatively small setups, however, his vision of expanding his home automation with many more IoT technologies has pushed him to create his own custom desktop application to keep track of the many devices that are connected to his smart speaker. The interviewee said that he would like the possibility to display the available commands and connections for each IoT device connected to his smart speaker. This highlights a known weakness in voice interfaces: it is hard to discover their capabilities.

**(3) Recovering from breakdowns**. When a breakdown occurred, participants tended use as the following strategies to recover from breakdowns: they either used their phone, or they oriented themselves towards the smart speaker and/or walked up to their smart speaker to repeat the command, though not always with success (e.g. see Figure 2). We also encountered a participant who experienced that his smart speaker turned on his smart light bulbs while he was away on vacation. On his smartphone, he was able to see that the lights were turned on although he had no means to do anything about it; the lights were on for a week. He was concerned about the conse-quences of connecting appliances to the smart speaker that would have more severe consequences when turned on for a whole week, such as an oven. This exemplifies control and intelligibility challenges that are perhaps inherent to a device such as a smart speaker that interfaces with various other IoT technologies and takes autonomous actions on the user's behalf: a person might not know that the system executed an action, why it did so, nor be able to intervene. Another participant shared a situation where the smart speaker from one month to the next did not recognize his request to turn on a light in a named room, despite it having worked previously. He assumed that something was updated, though he was unsure whether it was the Philips Hue lights or his Google Home smart speaker. This shows that users may find it difficult to understand where the problem lies when different software stacks are involved (e.g. Philips Hue, Google Home, Apple HomeKit). Another approach to provide control and intelligibility is for the smart speaker to offer alternative interpretations of the user's request. Three of the interviewees suggested this and the rest agreed, when we asked them about it, that it would be a good feature when presented as needed.

**(4) Shared use**. Finally, we noticed a commonality among the participants feeling fairly comfortable sharing their device with household members and visitors. While some were easing into using the smart speaker, others, usually visitors, showed a reluctance in using the devices. Some interviewees said that those guests were either not interested in the device, skeptical about privacy and the recording of voice interactions, or thought that this was a toy without any real use.

## OPEN QUESTIONS AND TOPICS FOR DISCUSSION AT THE WORKSHOP

### Physical Intelligibility

As observed in our study, the smartphone is the primary device many participants rely on when attempting to recover from a breakdown that goes beyond just repeating a request. Yet, many participants face and even walk over to the smart speaker before using the phone. This could provide an opportunity to explore explaining behaviours through the artifact's physical motion in relation to a person's presence and initial intended actions – what we call *physical intelligibility*. Physical motion has also been explored in prior work with other artifacts. Examples include Ju & Takayama's study on an approachable door that opened as to invite people in as they approached [9], and studies on the effect of motion on the perception of automatic systems [7,8]. Using physical intelligibility would provide non-frequent users, such as guests, new channels through which they can interact with the smart speaker, as well as provide regular users with alternative ways to interpret the device's state. When is it suitable to provide physical intelligibility to a user? In what way can physical intelligibility be provided to users? How can we draw on techniques from the field of robotics to design intelligible smart speakers? To what degree do people (owners and guests) need the smart speaker to be intelligible in its role as an agent and interface between other smart home and IoT devices?

## ABOUT THE AUTHORS

**Mirzel Avdic** is a PhD student in the Ubiquitous Computing and Interaction group at the Department of Computer Science at Aarhus University. He is researching how intelligibility and control can be alternatively and contextually designed to facilitate a better understanding and control of smart speakers and other IoT devices.

**Jo Vermeulen** is an Assistant Professor in the Ubiquitous Computing and Interaction group at the Department of Computer Science at Aarhus University. His research interests lie at the intersection of Human–Computer Interaction, Ubiquitous Computing and Information Visualization. He is particularly interested in designing interactive technologies that put people back in control of their digital environments.

ACKNOWLEDGMENTS

This project has received funding from the European Research Council (ERC) under the European Union's Horizon 2020 research and innovation programme (grant agreement No 740548).

### Alternative Interpretations

As our findings suggested, alternative interpretations could be useful. However, this also opens up further research questions due to the focus on voice interaction. How and when should a smart speaker provide alternative interpretations of a user's requests? In what way would such alternative interpretations influence the flow of "conversational" interactions between user and smart speaker and potentially break the metaphor of talking to an IPA? Using the smart speaker as a starting point to discuss alternative interpretations that such voice-controlled IoT devices have available, yet do not share with users, could be worthwhile discussing with respect to other (A-)IoT devices as well.

### Situational Physical Interaction with Smart Speakers

Our study shows that our participants prefer having the option to address a smart speaker and its connected IoT devices through other means than voice only, namely smartphone apps and the physical buttons on the speaker. While the smartphone is an obvious choice, it is worth questioning to which extent situational physical interactions (e.g. the participant who pressed the trigger button and whispered his request to the smart speaker) can be leveraged to provide alternative control mechanisms to users. How do we design for alternative smart speaker controls that go beyond voice interaction and smartphone applications, and when does it make sense to offer other means to recover from breakdowns? To what degree do people (owners and guests) need the smart speaker to be controllable in its role as an agent and interface with other smart home devices?